\documentstyle[emulateapj,psfig]{article}
\makeatletter

\newenvironment{inlinefigure}{%
\def\@captype{figure}%
\noindent\begin{minipage}{0.999\linewidth}\begin{center}}
{\end{center}\end{minipage}\smallskip}
\makeatother

\newcommand{\mum}{$\,\mu$m}

\newcommand{\coc}{CO(4--3) }

\newcommand{\co}{\rm CO}
\newcommand{\hh}{{\rm H}_2}

\newcommand{\msun}{\,M_{\odot}}

\newcommand{\Kkpspc}{\,\rm{K}\,\rm{km~s}^{-1}\,{\rm pc}^{2}}
\newcommand{\ssa}{SSA22-`Blob~1'}

\begin{document}
\title{Further multiwavelength observations of the SSA22
 Ly$\,\alpha$ emitting `blob'}
\author{S.C.\ Chapman,$\!$\altaffilmark{1}
D.\ Scott$\!$\altaffilmark{2}
R.A.\ Windhorst,$\!$\altaffilmark{3} D.T.\ Frayer,$\!$\altaffilmark{1}\\
C.Borys,$\!$\altaffilmark{1} G.F.\ Lewis,$\!$\altaffilmark{4} 
R.J.\ Ivison$\!$\altaffilmark{5}
}
\lefthead{Chapman et al.}

\altaffiltext{1}{Department of Physics, California Institute of Technology,
 MS 320-47, Pasadena, CA, 91125}
\altaffiltext{2}{Department of Physics and Astronomy, University of
 British Columbia, Vancouver, B.C., V6T1Z1~~Canada}
\altaffiltext{3}{Arizona State University, Dept.\ of Physics and Astronomy,
Tempe, AZ, 85287--1504}
\altaffiltext{4}{School of Physics, University of Sydney, NSW 2006~~Australia}
\altaffiltext{5}{Astronomy Technology Centre, Royal Observatory, Blackford 
Hill, Edinburgh EH9 3HJ, UK}

\slugcomment{Accepted in the Astrophysical Journal}

\begin{abstract}
We present new follow-up observations of the sub-mm luminous
Ly$\,\alpha$-emitting object in the SSA22 $z=3.09$ galaxy overdensity,
referred to as `Blob 1' by Steidel et al.~(2000).  In particular we
discuss high resolution {\sl Hubble Space Telescope\/} optical imaging,
Owens Valley Radio Observatory spectral imaging, Keck spectroscopy,
VLA 20\,cm radio continuum imaging, and {\sl Chandra\/} X-ray observations.
We also present a more complete analysis of the existing James Clerk Maxwell
Telescope sub-mm data.
We detect several optical continuum components which may be associated with 
the core of the submillimeter emitting region.
A radio source at the position of one of the HST components 
(22:17:25.94, +00:12:38.9) identifies it
as the likely counterpart to the submillimeter source.
We also tentatively detect the CO(4--3) molecular line, centered on the
radio position.
We use the CO(4--3) intensity to estimate a limit on the gas mass 
for the system.  The optical morphology of sources within the
Ly$\,\alpha$ cloud appears to be filamentary, while the optical source
identified with the radio source has
a dense knot which may be an AGN or compact starburst.
We obtain a Keck-LRIS spectrum of this object, despite its 
faintness ($R=26.8$). The spectrum reveals weak Ly$\alpha$ emission, but
no other obvious features, suggesting that the source is not an energetic
AGN (or that it is extremely obscured).
We use non-detections in deep {\sl Chandra\/} X-ray images
to constrain the nature of the `Blob'.
Although conclusive evidence regarding the nature of the object remains hard
to obtain at this redshift, the evidence presented here is at least consistent
with a dust-obscured AGN surrounded by a starburst situated at the heart of
this giant Ly$\,\alpha$ cloud.
\end{abstract}

\keywords{cosmology: observations --
galaxies: evolution -- galaxies: formation -- galaxies: starburst}

\section{Introduction}
\label{secintro}

Deep surveys of the submillimeter (sub-mm) sky using the Submillimeter
Common-User Bolometer Array (SCUBA) on the James Clerk Maxwell
Telescope have uncovered a population of distant dust-rich galaxies
(see Blain et al.~2002 and references therein).  Based on the 
radio/sub-mm indices, optical colors, and the spectroscopic
identifications for $>60$ submillimeter galaxies, 
the majority of these systems are 
thought to lie at redshifts of
$z\sim 1$--4 (e.g.~Smail et al.~2002, Chapman et al.~2003a, 2004).  

Identifying the counterparts of sub-mm sources at other wavelengths has proven
difficult, due to the large beamsize of sub-mm instruments, and the
inherent faintness of the sources at all shorter wavelengths.
The radio regime has emerged as an efficient means to pin-point the sub-mm
sources (through the FIR/radio relation, e.g.\ Helou et al.~1985; Condon 1992), however
when radio emission cannot be detected, the only recourse is to use
millimeter interferometry to attempt to localize the source.

The first well-studied sub-mm
system SMM\,J02399--0136 (hereafter SMM\,J02399) at $z=2.8$ was shown to
contain both an AGN (Ivison et al.~1998) and a massive reservoir of
molecular gas thought to be fueling a starburst (Frayer et al.~1998).
This scenario is increasingly becoming  
the conventional paradigm for the sub-mm population.
This is perhaps unsurprising, given that both AGN and starbursts are
thought to be triggered by galaxy interactions (e.g.~Sanders et al.~1988,
Archibald et al.~2002).
With the proliferation of spectroscopic redshifts for sub-mm sources, many
showing AGN features (Chapman et al.~2003a, 2004), 
detecting molecular CO gas has become 
almost routine (Frayer et al.~1998, 1999; Neri et al.~2003; Greve et al.
2004).

Sub-mm sources identified in the optical are typically suggestive
of mergers in progress, based mainly on ground-based images of disturbed,
multiple component structures (e.g., Smail et al.~2002, Ivison et al.~2002). 
However, imaging at {\sl HST\/} resolution exists for very few
sub-mm sources, making the detailed morphological study of sub-mm galaxies
difficult. {\sl HST\/} imaged examples of 12 robustly identified SCUBA
galaxies reveal the ground-based structure to be often a complex of 
many smaller fragments (Chapman et al.~2003b). 
The fragmented, merger morphology of SCUBA galaxies observed
by {\sl HST\/} in the SA13 deep field (Sato et al.~2002), and of
the SCUBA luminous Lyman-break galaxy, Westphal-MMD11 
(Chapman et al.~2002) seem representative of the sub-mm population. 

One of the intrinsically brightest sub-mm sources yet discovered is
the \ssa\ (Chapman et al.~2001), lying in the overdense
core of a possible proto-cluster of galaxies at $z=3.09$
(Steidel et al.~2000).
The nature of this source remains enigmatic, despite the existing
multi-wavelength detections and deep ground-based imagery and spectroscopy.
The extent to which this object is representative of high-$z$ sub-mm detections
is, as yet, quite unclear.  It was targeted with SCUBA only after the
extended Ly$\,\alpha$ emission was already known.  
However, highly clustered environments may be typical of 
many blank field sub-mm galaxies (Blain et al.~2004).
In addition,  
extended gaseous haloes have been detected around several other 
sub-mm galaxies:
SMM\,J02399 (Ivison et al.~1998), 4C41.17 (Ivison et al.~2000b), 
SMMJ\,17142, in the field of radio galaxy 53W002 (Smail et al.~2003a) 
and SMM\,J16034 (Smail et al.~2003b).
Whether representative or unique, the combination of extended Ly$\,\alpha$
emission and the presence of large amounts of dust (which is effective at
destroying Ly$\,\alpha$) is surprising and merits further study.
Some ideas for what may be going on include: a superwind from an
extreme starburst (Taniguchi, Shioya \& Kakazu~2001; Ohyama et al.~2003);
cooling radiation in a forming galaxy halo (Fardal et al.~2001);
or some contribution from the Sunyaev-Zel'dovich effect increment.
Chance superposition or the effects of gravitational lensing can also
complicate the interpretation.

In this paper we present new {\it Hubble Space Telescope\/} optical 
imaging, Keck-LRIS spectroscopy, 
as well as
OVRO CO(4--3) measurements and a re-analysis of the available SCUBA,
VLA-radio, and {\sl Chandra\/} X-ray
data.  We use this combination of multi-wavelength data
to localize the sub-mm emission within the optical/near-IR images and
to discuss the spectral energy distribution (SED) of the `Blob'.
All calculations assume a flat $\Lambda$CDM cosmology with
$\Omega_\Lambda=0.7$ and $H_0=65$\,km\,s$^{-1}$\,Mpc$^{-1}$, so that 1\,arcsec 
corresponds to 8.2\,kpc at $z=3.09$,
where the luminosity distance is 28360\,Mpc.

%
%
\begin{figure*}[htb]
\centerline{
\psfig{figure=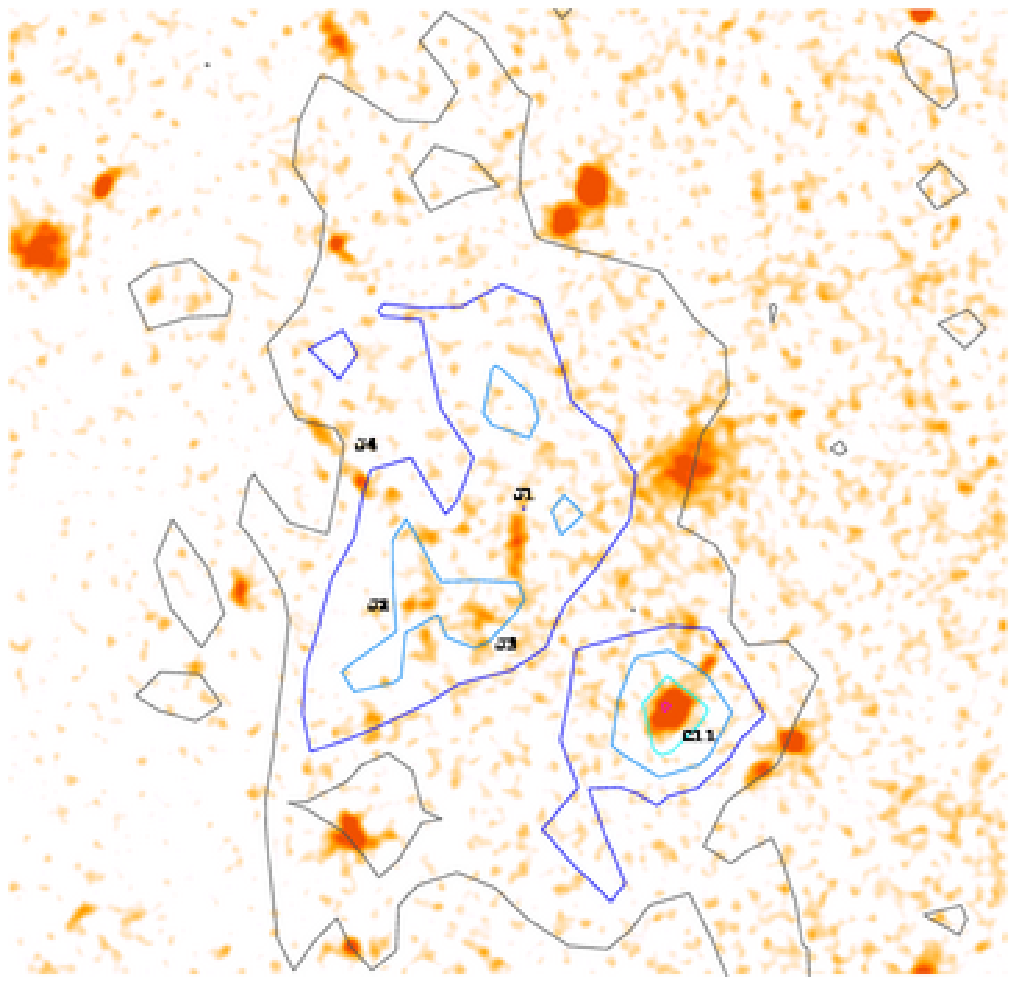,angle=0,width=3.6in}
\psfig{figure=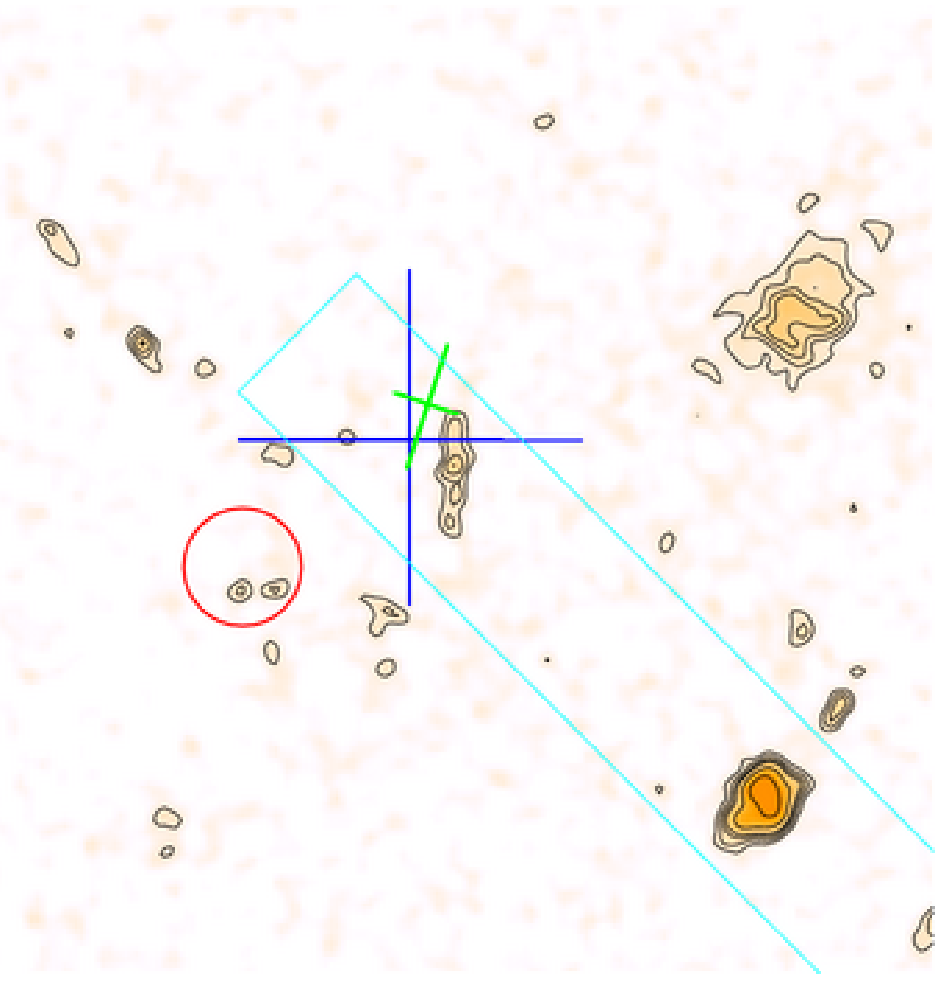,angle=0,width=3.5in}
}
\figurenum{1}
\caption{
The \ssa\ region (North is up, East is left)
observed with {\sl HST\/}-STIS 50CCD.
The left panel greyscale image is $18\times18$ arcsec, with contours overlaid 
showing the extended Ly$\,\alpha$ cloud obtained
from the William Herschel Telescope, SAURON, integral field spectrograph image
(Bower et al.~2004).
The {\sl HST\/} source J1, detected in our VLA-radio image, is labeled,
along with Ly$\,\alpha$ knot J2, and an additional component J3, any 
of which might be related to the SCUBA and OVRO centroids ($1\sigma$
centroiding plus pointing errors indicated by the
large and small crosses, respectively).
An extended linear feature to the NE is labeled J4, and
the Lyman Break Galaxy LBG-C11 is also indicated.  
The right panel image shows a $9\times9$ arcsec zoomed greyscale, with
contours of the {\sl HST\/}-STIS image starting with 3$\sigma$ and increasing
by 1$\sigma$). 
The $K_{\rm s}=21.5$ source detected by Steidel et al.\ (2000) using Keck/NIRC
is identified as a circle, corresponding to our J2 (compare their Figure~7).
The Keck-LRIS slitlet placement over J1 and C11 is also overlaid.
The bright object in the north-west of the image has $U,g,R,I,K$ colors 
which make it inconsistent with a $z\sim3.09$ galaxy, and is likely 
at much lower redshift.
}
\label{fig1}
\addtolength{\baselineskip}{10pt}
\end{figure*}
\section{Observations and Analysis}
\subsection{HST-Visible observations}

{\sl HST\/} imaging was obtained with the
Space Telescope Imaging Spectrograph (STIS).
Three orbits of {\it LOW SKY} integration time were 
split between six exposures, using the 50CCD-clear filter,
providing 7020\,sec.
Pipeline processed frames were calibrated, aligned, and cosmic ray rejected,
using standard {\sc IRAF/STSDAS} routines. 
The pixel size in the STIS image is 0.0508\arcsec. The sensitivity
limit reached is $27.6$ mag (5$\sigma$), 
corresponding to $R\sim28.6$ for a point
source with a late-type spiral galaxy SED.
The 50CCD-clear filter is roughly a Gaussian with 1840\AA\ halfwidth
and an effective wavelength of 5733.3\AA. 
We refer to the associated AB magnitude as $R'(573)$ hereafter.
The STIS image is presented in Fig.~1 with a Ly$\,\alpha$ outline overlaid
(Bower, private communication).
The relative astrometry was carried out by matching all
bright sources, providing a 
relative positional error of 0.28\arcsec.

Whereas no optically detected continuum sources were present within the
core of the Ly$\,\alpha$ cloud from the $R\sim26$ ground based imagery
in Steidel et al.\ (2000), we find
several compact and distorted sources in the deep {\sl HST\/}-STIS imagery
(Fig.~1), labeled J1, J2, J3 and J4.
However, comparison of our {\sl HST\/} image with the Ly$\,\alpha$ contours
from Steidel et al.\ (2000) and Bower et al.\ (2004), 
suggests that these {\sl HST\/} sources do not always
trace the density of Ly$\,\alpha$.  For example, there was no narrow-band
knot at the position of J1 or J4, although J2 and J3 sit near Ly$\,\alpha$
peaks.
The $R'(573)$ 1\arcsec
aperture magnitudes for the different components using the SExtractor
package (Bertin \& Arnouts 1996) are: J1$\,{=}\,26.82$; J2$\,{=}\,27.12$;
J3$\,{=}\,27.65$; and J4$\,{=}\,26.44$.
Only the J2 component is detected in the $K$ band ($K_s=21.5$, Steidel
et al.\ 2000).

%
%
\begin{figure*}[htb]
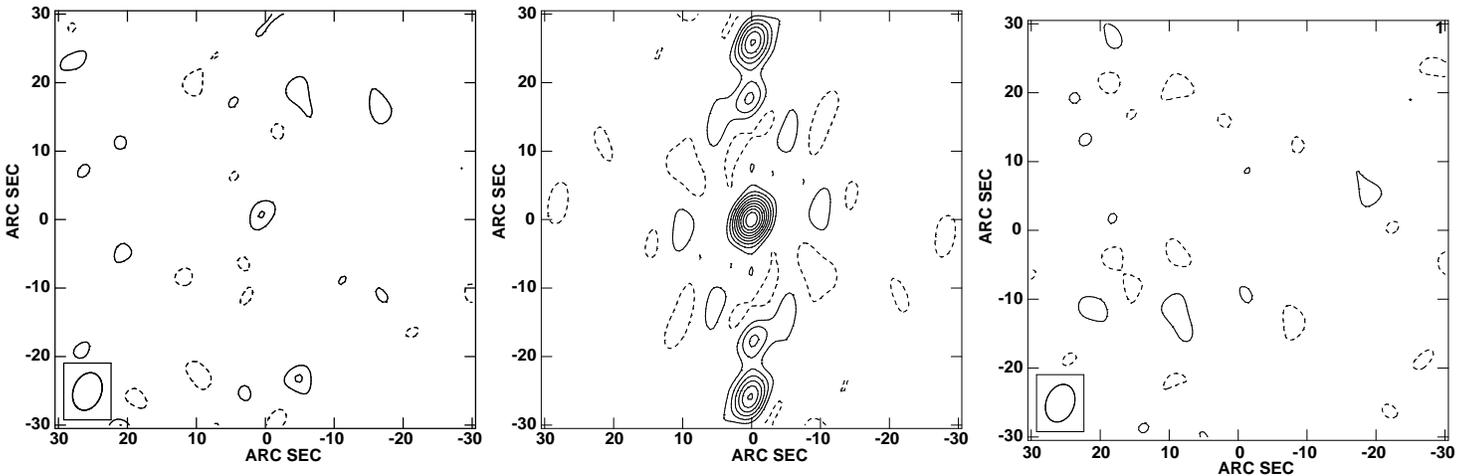

\centerline{
\psfig{figure=sa22map.ps,angle=0,width=2.5in}
\psfig{figure=SA22_BEAM.PS,angle=0,width=2.5in}
\psfig{figure=SA22_1.PS,angle=0,width=2.5in}
}
\vspace{5pt}
\figurenum{2}
\caption{
OVRO spectral imaging.  The left panel shows the {\it dirty beam},
higher frequency half-band
containing the marginally detected $400\,{\rm km}\,{\rm s}^{-1}$ CO(4--3)
line near the centroid of the SCUBA beam.
Excess power near the positions of the beam sidelobes is consistent with
our tentative detection.
The OVRO beam is shown as an inset, while the dirty beam pattern
sidelobes are shown in the middle panel.  The right panel shows
the lower frequency half-band dirty beam map, 
revealing no source $>3\sigma$ within the
central 20\arcsec$\times$20\arcsec. Contours are 2$\sigma$, $3\sigma$,
etc., times the rms ($0.8\,{\rm Jy}\,{\rm km}\,{\rm s}^{-1}{\rm beam}^{-1}$)
with negative contours dashed.
}
\label{fig2} 
\end{figure*}

\subsection{OVRO CO(4--3) observations}
\label{secovro}

\ssa\ was observed using the Owens Valley Millimeter Array over 
8 hour tracks scheduled through 
March and June of 2001.  A total of 28 hours of high quality
integration time on-source was obtained in good winter weather conditions,
in two configurations of six
$10.4\,$m telescopes.  The phase center for the CO observations was the
position of the brightest near-IR component of the \ssa\ cloud
identified as an $(R\,{-}\,K)\,{>}\,6$
Extremely Red Object (ERO) in Steidel et al.~(2000), which we now refer to
as J2: 
$\alpha$(J2000)=$22^{\rm h} 17^{\rm m}25\fs93$,
$\delta$(J2000)=$+00\arcdeg 12\arcmin 37\farcs6$.
The \coc line was observed using a digital correlator configured with
$112\times4\,$MHz channels, centered on $112.390\,$GHz in the lower
side-band, corresponding to \coc emission at a redshift of
$z=3.102$. 

Choosing an observing frequency for molecular lines
is difficult when only the Ly$\,\alpha$ emission line redshift is known.
The closest bright Lyman Break Galaxy in the proto-cluster is C11, with
$z=3.1080$ for Ly$\,\alpha$. The redshift for the Ly$\,\alpha$
knot $\sim$7\arcsec\ farther along the slit, and associated
with the $K=21.5$ source, is identical within
the errors. Using the interstellar absorption lines for LBG-C11
gives $z=3.0964$. From near-IR spectroscopy, 
the systemic redshift is typically half-way between the
Ly$\,\alpha$ and absorption redshifts (Pettini et al.~2001,
Adelberger et al.~2003), suggesting $z\simeq3.102$.
However, there is no detected continuum in the spectrum of J2
and it is not clear that the same estimate should necessarily apply.
Hence the uncertainty in the systemic redshift is probably around 
$400\,{\rm km}\,{\rm s}^{-1}$.
The Keck spectrum of \ssa\ (see below) shows weak Ly\,$\alpha$ in
emission, $220\,{\rm km}\,{\rm s}^{-1}$ to the red
of LBG-C11 (or at a redshift $z=3.111$). 
The low S/N of the spectrum, coupled with uncertainties in the systemic 
redshift suggest that this new information would not improve the estimated
redshift for the CO measurement.

Typical single-sideband system
temperatures were approximately 400--$500\,$K, corrected for telescope
losses and the atmosphere.  In addition to the CO line data, we recorded
the $3\,$mm continuum data with a $1\,$GHz bandwidth for both the upper
(line-free, centered on $112.9975\,$GHz) and lower sidebands.  The nearby
quasar 2213+035 was observed every 25 minutes for gain and phase
calibration.  Absolute flux calibration was determined from observations
of Uranus, Neptune, and 3C\,273.  The absolute calibration uncertainty
for the data is approximately 15\%.
The 95\% confidence upper-limit for
the continuum emission is S$_{\nu}$(3\,mm) $< 0.9$\,mJy.
This is insufficient to detect the expected thermal dust emission discovered by
SCUBA, assuming a dust spectrum $S_{\nu} \propto \nu^{3.5}$.

Fig.~2 shows the \coc\ spectral map for the `Blob'.  The CO line is tentatively
detected at a redshift offset $-100\,$MHz from our central frequency,
corresponding to $270\,{\rm km}\,{\rm s}^{-1}$ shift to the red.
The line width is estimated at $400\,{\rm km}\,{\rm s}^{-1}$ FWZI.
We achieved a $3.2\sigma$ detection for the peak
in the integrated \coc\ map (upper frequency half-band map).  
The CO position
appears offset to the north by 2\arcsec\ from 
the ERO/Ly$\,\alpha$ peak position. 
The OVRO beamsize (4\arcsec\ by 6\arcsec) 
is shown as inset, while the beam pattern
sidelobes are shown in the middle panel.
An apparent excess power at the sidelobe positions of our \ssa\ CO map 
is consistent with a tentative detection.
The CO map shows no obvious evidence for
extended emission. 
The lower frequency half-band map is also shown in Fig.~2 (right panel),
finding no sources 
above 3$\sigma$, and no sources $>2\sigma$ within 10\arcsec\ of the center.
We caution that the noise from the interferometer is unlikely to be Gaussian,
and the 3$\sigma$ noise limit could be an underestimate.

We interpret the OVRO data as a marginal detection of the CO line, of
width around $400\,{\rm km}\,{\rm s}^{-1}$ -- the Gaussian width we will 
assume for the analysis of the molecular gas properties in \S3.4.

\subsection{VLA Radio Observations}
\label{secvla}

VLA observations were obtained in the `A' (36\,hours) and `B' (12\,hours)
configuration at 1.4\,GHz.
The map reaches an r.m.s. sensitivity of 8.5\,$\mu$Jy near the phase center
of the map. Reductions and details are described elsewhere
(R.J.\ Ivison, in preparation).
We used these observations to
search for a radio counterpart to the sub-mm source in order to
pin-point the location of the far-IR emission (Fig.~\ref{vla}).
A 4.4$\sigma$ radio source is detected at the optical position of J1,
22:17:25.94 +00:12:38.9 in the radio FK5 grid,
lying well within the SCUBA beam.
The primary beam corrected flux of the radio source is 44.4$\pm$10.1\,$\mu$Jy.
The detection is significant above the non-Gaussian
{\it phase} noise in the interferometric data (Richards 2000),
as the only peak $>4\sigma$ (positive or negative)
within the square arcmin region of Fig.~\ref{vla}
other than the neighboring bright radio source 20\arcsec\ to the West.
At $z=3.09$ this corresponds to a luminosity of
$\nu L_\nu = 3.0\times10^8\,{\rm L}_\odot$.
The neighboring western
radio galaxy (188\,$\mu$Jy peak, 242\,$\mu$Jy integrated) 
identifies a second SCUBA detection in the field
(Fig.~\ref{scuba}) at 22:17:24.682 +00:12:42.02.

%
%
\begin{inlinefigure}
\vspace{6pt}
\psfig{figure=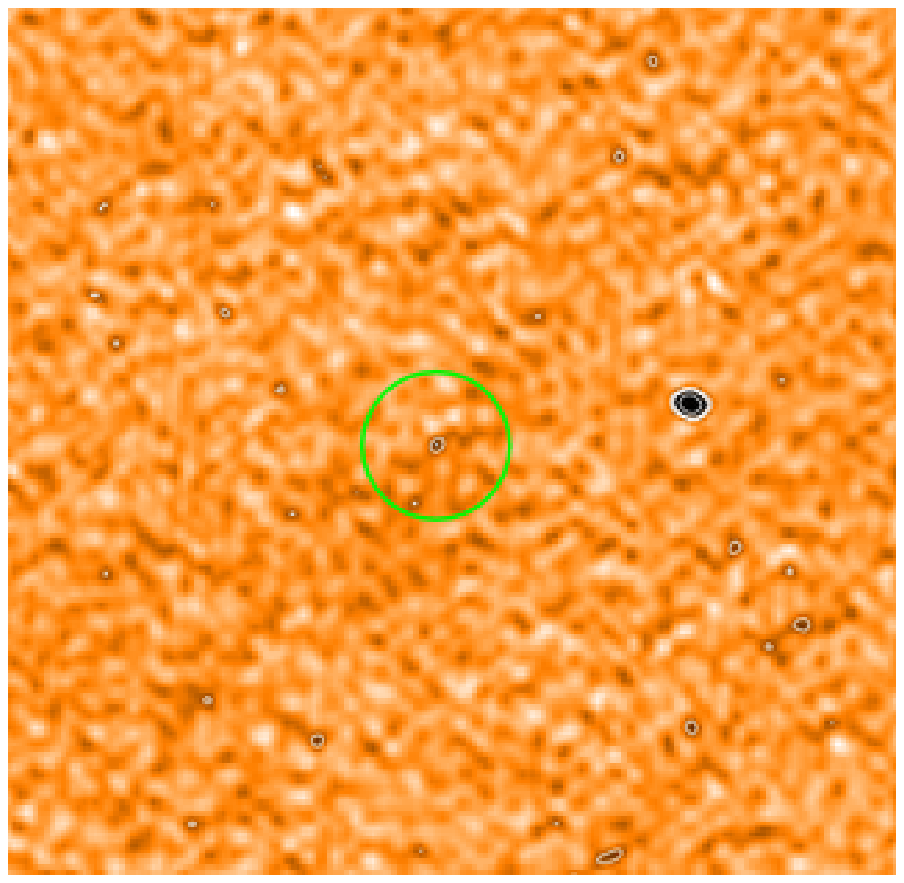,angle=0,width=3.5in}
\vspace{6pt}
\figurenum{3}
\caption{
A 1~arcmin$^2$ 
20\,cm radio map of \ssa.
The contours represent 3, 4, 5, 10, and 20 times the rms noise of 10.1\,$\mu$Jy.
The \ssa\ source (J1) itself has a flux density of 44.4\,$\mu$Jy.
A neighboring bright source (189\,$\mu$Jy) 
to the West identifies a weaker submm source
seen in Fig.~\ref{scuba}.
The circle is 10\arcsec\ diameter, and depicts the position of J1.
}
\label{vla}
\addtolength{\baselineskip}{10pt}
\end{inlinefigure}

\subsection{Chandra X-ray Observations}
\label{secxray}

A 70\,ksec {\sl Chandra\/} ACIS-I integration was taken, centered on the
SSA22 field. The data were procured from the {\sl Chandra\/} archive
and searched in the vicinity of \ssa.  
No significant counts were recorded over the `Blob-1' region, where
a $3\sigma$ limit of 85 counts was achieved (although the adjacent
candidate sub-mm source $\sim$30\arcsec\
to the West has a weak X-ray counterpart).
This limit on extended emission over the Ly\,$\alpha$ `Blob' 
is not very restrictive, with 2.8$\times10^{-17} {\rm W}\,{\rm m}^{-2}$
from the 0.2--$10\,$keV band
for a 4\,keV thermal spectrum at $z=3.1$. In our adopted cosmology, this
translates to $L_{\rm x} < 2.1\times 10^{38} {\rm W}$,
which corresponds only to the upper mass range of rich galaxy clusters
(e.g.\ Edge 2003).
The point source limit for `J1' however, is much more restrictive 
($L_{\rm x} < 1.3\times 10^{37} {\rm W}$), well within the regime of
local low-luminosity AGN, similar to Seyfert galaxies.

\subsection{SCUBA Sub-mm Observations}
\label{secscuba}
We have also obtained new SCUBA observations of the `Blob'.
These jiggle-map mode observations were taken during April\ 2001.
In addition, archival observations (PI: A.~Barger) 
covering part of the region were retrieved
and combined with the new data.
The final combined map detects the main
\ssa\ source with $S_{\rm 850 \mu m}=16.8\pm2.9$\,mJy, 
$S_{\rm 450 \mu m}=45.1\pm15.5$\,mJy.
The separate data sets yield consistent results.
We checked that there is no indication that the core of the 
source is extended,
although there seems to be extent in the emission towards the west.
This may simply be confusion with 
other apparent sources more than a beam-width away in the 
SCUBA $850\,\mu$m image.
This significant ($3.8\sigma$)
extention of the central source, peaking 21\arcsec\ to the
West of \ssa, is identified with a bright (188\,$\mu$Jy peak) radio source.

The range of structures present in this field, and the multiple chop  
throw angles used for the different subsets of the data, 
may give rise to artificial structures caused by off-beams.
The S/N of the image does not merit a detailed deconvolution.
The combined SCUBA image is shown in Fig.~\ref{scuba}.

%
%
\begin{inlinefigure}
\vspace{6pt}
\psfig{figure=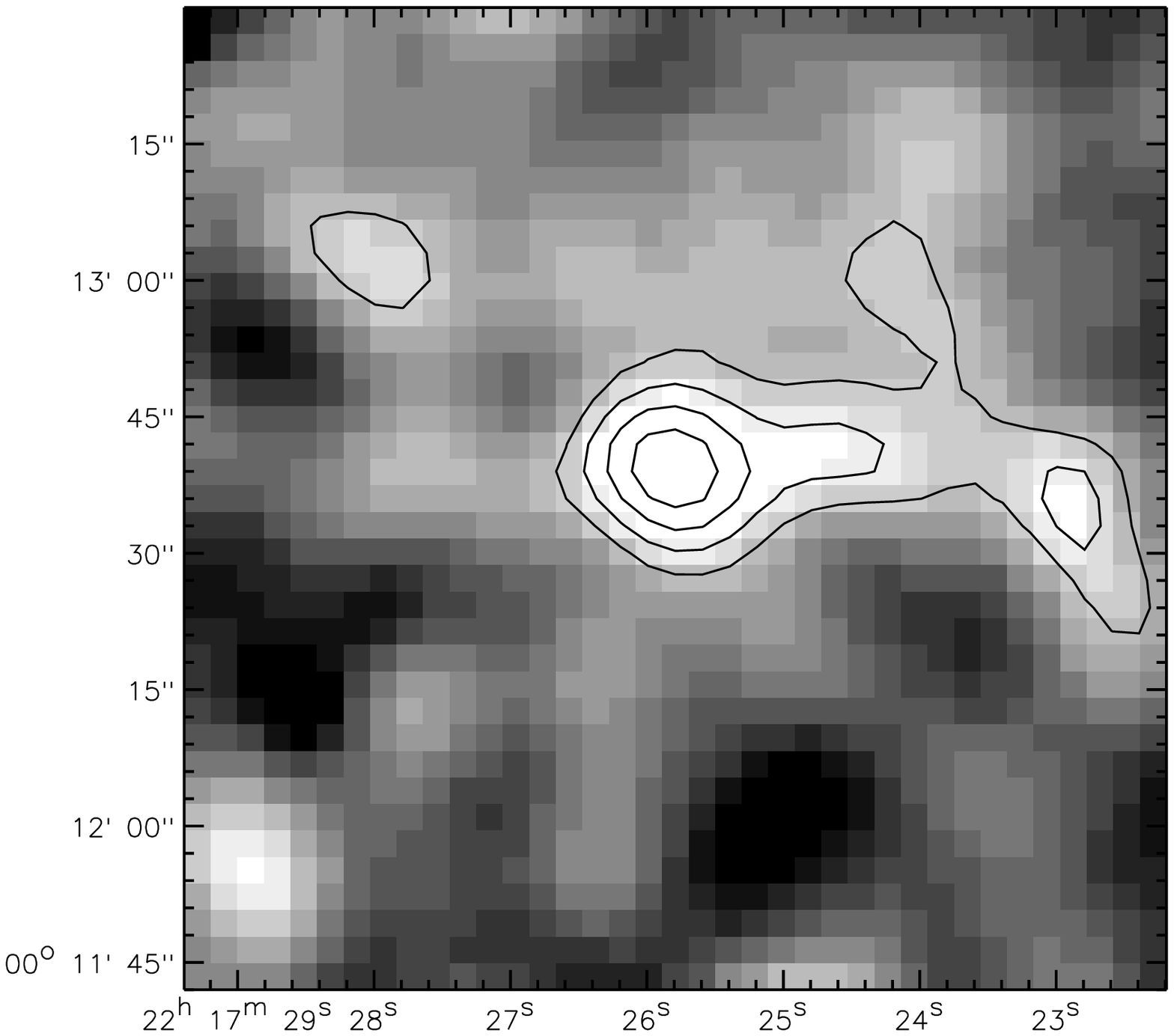,angle=0,width=3.5in}
\vspace{6pt} 
\figurenum{4}
\caption{
$850\,\mu$m SCUBA map of \ssa\ using data combined from several observing
runs in jiggle-mapping mode.
The contours represent 2, 3, 4, and 5 times the rms noise of 2.9\,mJy.
The \ssa\ source itself has a flux density of 16.8\,mJy. 
A significant extention of the central source, peaking 21\arcsec\ to the
West of \ssa, is identified with a bright (189\,$\mu$Jy) radio source
at 22:17:24.682, +00:12:42.02.
Lower significance
sources are present at the periphery of the SCUBA image, notably a 3.6$\sigma$
peak 35\arcsec\ to the West of \ssa\ and a 2.9$\sigma$ peak 30\arcsec\ to the
North-east.}
\label{scuba}
\addtolength{\baselineskip}{10pt}
\end{inlinefigure}

\subsection{Keck-LRIS spectroscopy of `Blob~1'}
Keck LRIS spectroscopy of the `Blob' {\sl HST\/} identified source (J1)
was taken
using the 400 lines/mm grism, providing a spectral resolution of $\sim$10\AA\
(Fig.~5). 
A slitlet was positioned on the source J1,
with a position angle of $-140^\circ$, allowing the LBG, C11, lying 7\arcsec\ 
southwest of J1 to also be aligned on the slit
(Fig.~5, offset brighter spectrum).
This facilitated the 1-dimensional extraction of the extremely faint $R\sim27$
continuum of J1.
The spectrum shows detected Ly\,$\alpha$ emission, and possibly low significance
features commonly found in LBG spectra (Shapley et al.~2003).

Ohyama et al.~(2003) also recently presented optical spectroscopy of 
the `Blob' using SUBARU-FOCUS. While their deep $R$-band image detects
the HST sources (J1, J2, J3, J4) in 0.5\arcsec\ seeing, 
their spectrograph slit position misses all these components.

The weak  Ly\,$\alpha$ line in J1 shows a rest equivalent width of $18$\AA\
and an essentially unresolved width (${<}\,200\,{\rm km}\,{\rm s}^{-1}$).
The lack of detectable high ionization lines (CIV, SIV), and the
relatively narrow Ly\,$\alpha$ line suggest that an energetic AGN
is unlikely to be present (or else it is highly obscured).
There is a redshift offset to the red 
in J1 from the Ly\,$\alpha$ peak of the LBG C11,
corresponding to a shift of $220\,{\rm km}\,{\rm s}^{-1}$.
However, large offsets from systemic velocities
due to stellar winds are likely (Adelberger et al.\ 2003), and we cannot
interpret this offset as physical displacement. 
We note that the redshift of the peak in the tentative CO measurement is
very close to the J1 Ly\,$\alpha$ peak ($50\,{\rm km}\,{\rm s}^{-1}$
redwards of the J1 Ly\,$\alpha$).

\section{Results}

\subsection{Spectral Energy Distribution}

A spectral energy distribution (SED) for the `Blob' with all the new data
points and upper limits is shown in Fig.~6.  In particular we have indicated
the OVRO continuum limit as well as the marginal line detection.
A range of dust temperatures are reflected by the four overlaid SED
templates (25, 29, 34 and 50\,K, for a dust emissivity $\beta\,{\simeq}\,1.5$),
with the 34\,K SED fitting the sub-mm
points the best. The direct extrapolation of this SED to the
radio using the far-IR/radio correlation (Helou et al.\ 1985) is
consistent within the error bars of our radio detection.

The various optically detected fragments
are fainter than 80\% of SCUBA galaxy identifications (Chapman et al.\ 2003b), 
however the components vary considerably in their $R-K$ color.
A steep ($\alpha=-2.5$) power-law is required for consistency with the
$R-K$ limit on `J1'.

\subsection{The location of the sub-mm emission}

Since the `Blob' appears to be a unique object, we would like to 
relate the emitting regions at various wavelengths in order to study
its physics.
The radio source is coincident with the $HST$ source, J1, to within the
astrometric errors aligning the optical/radio frames ($\sim$0.4\arcsec).
The relative astrometry between the OVRO/SCUBA data and the {\sl HST\/}
frame is achieved by mapping the SSA22 radio sources onto a deep
$I$-band image and matching the {\sl HST\/} sources to this frame.

%
%
\begin{inlinefigure}
\centerline{
\psfig{figure=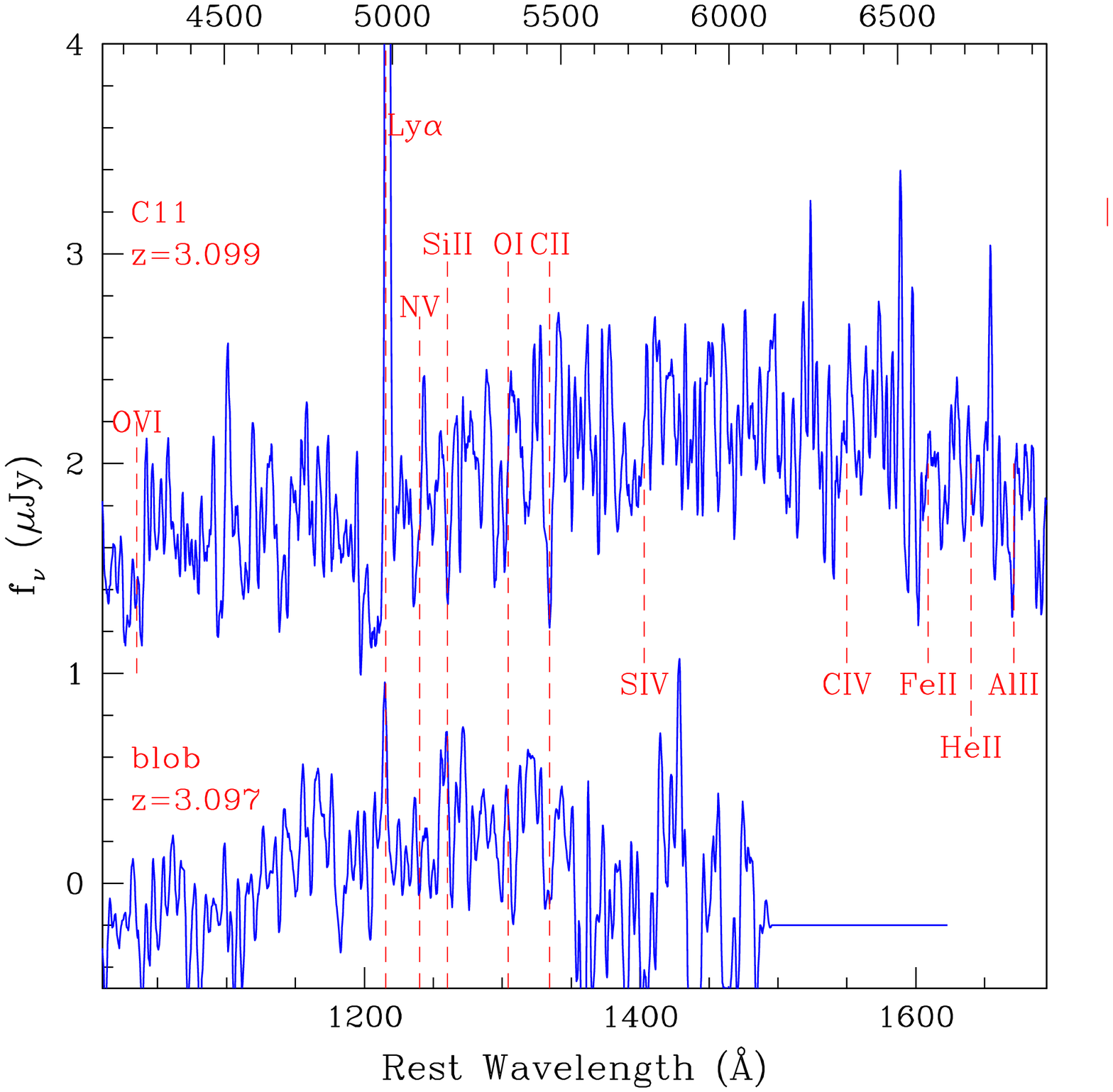,angle=0,width=3.5in}
}
\vspace{5pt}
\figurenum{5}
\caption{
Keck LRIS spectroscopy of the `Blob' {\sl HST\/} identified source J1
(lower spectrum), 
revealing a detection of Ly$\alpha$ emission, and possible lower significance
features commonly found in LBG spectra (Shapley et al.~2003).
The LBG, C11, lying 7\arcsec\ to the southwest
of the J1 was also aligned on the slit (shown above,
offset by 2$\,\mu$Jy for clarity)
facilitating the 1D extraction of the extremely faint $R\sim27$ 
continuum.
}
\label{fig5}
\end{inlinefigure}

The r.m.s.~variation in pointing errors with the JCMT/SCUBA observations
were typically $\simeq2.5\arcsec$ throughout these observations, 
as measured by offsets to pointing sources, and this
dominates the centroiding of the sub-mm source.
Assuming that we can pin-point a source to the FWHM divided by 2.35 times the
signal-to-noise ratio, and adding the errors in quadrature, the
estimated error in the SCUBA position is about 3\arcsec.
This error is overlaid on the {\sl HST\/} image with the large cross in Fig.~1.
{}From the tentative CO(4--3) detection from the OVRO peak, we obtain a
centroiding error $\simeq1.0\times$0.6\arcsec, dominating the positional
uncertainty of the phase reference. 
This is shown with the smaller cross in Fig.~1.

%
\begin{inlinefigure}
\vspace{26pt}
\psfig{figure=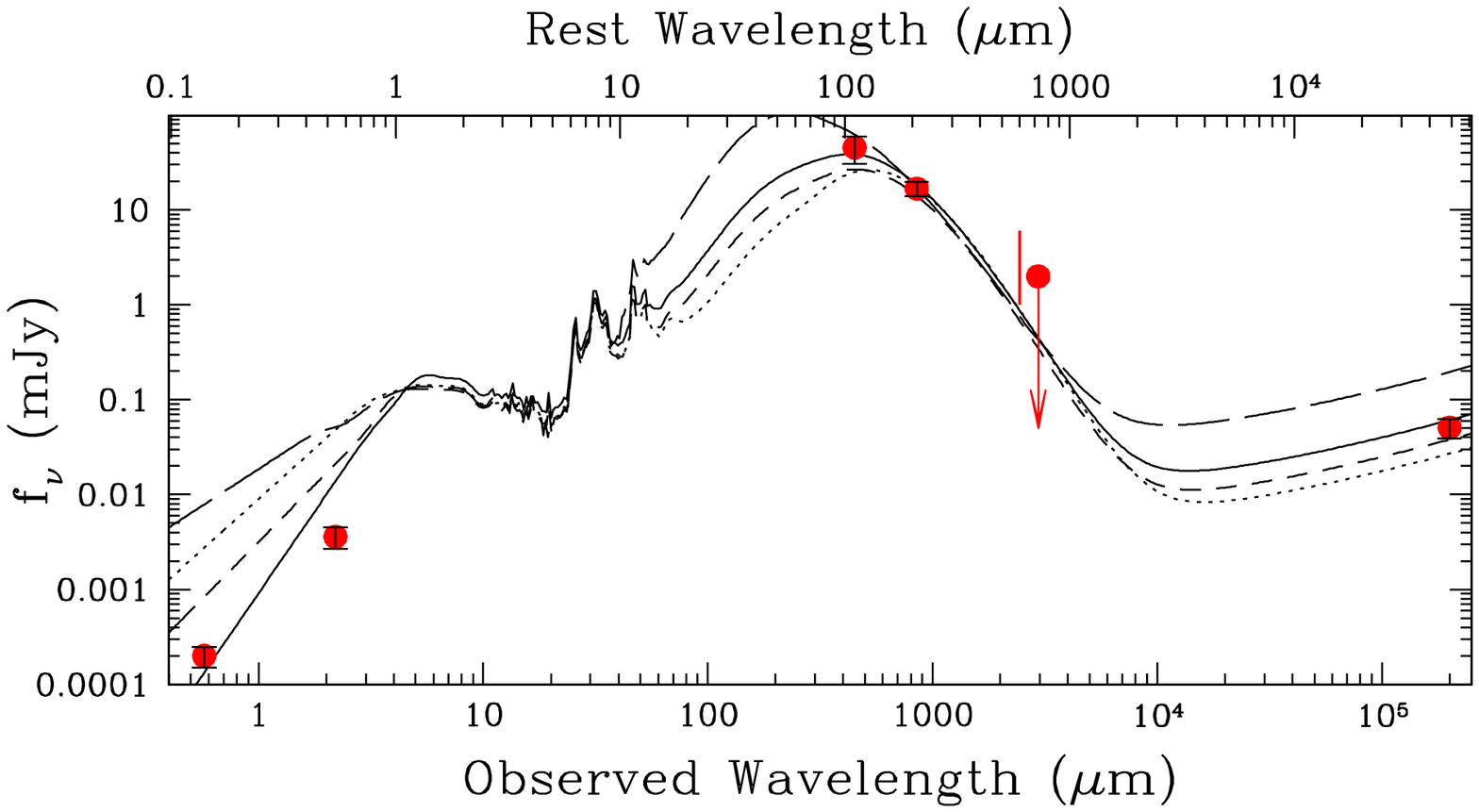,angle=0,width=3.5in}
\vspace{6pt} 
\figurenum{6}
\caption{
Spectral Energy Distribution of \ssa~showing the new data points presented
in this paper at $20\,$cm, $3\,$mm, 850\mum, 450\mum, $K$-, and
$R^\prime(573)$-band.  The OVRO data are shown both for the continuum
upper limit and the marginal CO line detection, slightly offset for clarity.
The curves show three SEDs from the library of Dale et al.\ (2001), with
dust temperatures of 25\,K, 29\,K, 34\,K, and 50\,K, normalized to the 850\mum\ point.
}
\label{fig6}
\addtolength{\baselineskip}{10pt}
\end{inlinefigure}

The most likely association of the OVRO/SCUBA source is therefore with the 
51\,$\mu$Jy radio source corresponding to
the brightest linear feature in the {\sl HST\/} image, denoted J1 in Fig.~1.
We can therefore likely rule out J2 or J3 as contributing significantly
to the submillimeter emission.
Note that the component J2 is the ERO source described in Steidel et al.
(2000), which we now measure to have $(R\,{-}\,K)=6.5$.
As this is {\it not\/} the source of the sub-mm emission,
it is at odds with findings from other sub-mm systems, that nearby
$(R\,{-}\,K)>6$ ERO components, often seem to be the sub-mm source
(Smail et al.\ 2000; Frayer et al.\ 2000, 2003, 2004; Webb et al.\ 2004).
J1 has an $(R\,{-}\,K)<4.3$, to the 3$\sigma$ limit of our $K$-band data.

\subsection{Linear Features in the HST Image}
We also note the striking
linear structure between
J1 and J2 that sticks out towards the NE -- labeled J4 in Fig.~1).  
At approximately 5 arcsec in length,
this is quite unusually long and linear compared with chain-like galaxies
(e.g.~Cowie et al.~1995) in similar {\sl HST\/} images.

Could this linear object-chain be jet-induced star-formation
from the hidden AGN in J1 (or perhaps from another AGN in J2 or even in C11?).
This is similar to what was suggested might
be the case in 53W002 at $z=2.39$ (Windhorst, Keel \& Pascarelle 1998).
If so, it is
curious that we see no jet-like radio or X-ray source, in fact no radio or
X-ray source is seen at all at the position of J4 (although a weak radio 
source does identify the J1 optical feature).  
With little detailed information, we can offer little more than speculation.
The optical `jet' may not dim as much as it would in the radio or X-ray,
since here we may be possibly looking at a string of star-bursting knots
-- putatively all induced by the AGN jet -- but each essentially unresolved
by STIS.

The unresolved peak within the linear structure of J1 is at least
suggestive of an AGN, although without other diagnostics, we cannot
rule out a compact starburst.
The apparent limit on the X-ray emission is not sufficient to rule out an
obscured AGN.
Spatially resolved spectroscopy along the axis of the linear feature
could in principle differentiate emission mechanisms, however, the 
fragment is extremely faint in the optical and our present Keck-LRIS
spectrum likely represents the best practical quality achievable with a
10\,m class telescope.

\subsection{Molecular gas}

The limit on the integrated \coc line flux from the marginal
detection of the `Blob'
represents $S(\co)\,<\,2.5\,{\rm Jy}\,{\rm km}\,{\rm s}^{-1}$.  
No adjustment has been made to account for the continuum level since it
appears to be negligible.
The observed \coc line flux (peaking at $\sim 5$\,mJy) 
implies an intrinsic CO line luminosity
$L^{\prime}(\co) = 7.5\times 10^{10} \Kkpspc$
(see the formulae in Solomon, Downes, \& Radford 1992).  The CO luminosity
is related to the mass of molecular gas (including He) by
$M(\hh)/L^{\prime}(\co) = \alpha$, with the value for $\alpha$ expected
to be $\simeq 1 \msun(\Kkpspc)^{-1}$, consistent with that estimated
for local ultraluminous infrared galaxies (ULIGs; Solomon et al.\ 1997). 
We then adopt a correction for the 
excitation CO(4--3):CO(1--0) brightness ratio of $\simeq0.5$ typically
observed in starbursts (Devereux et al.~1994), yielding 
$\alpha=2 \msun(\Kkpspc)^{-1}$.
The inferred limit on the 
molecular gas mass of the `Blob' is $2 \times 10^{10} M_{\sun}$,
which is consistent with that of the most massive low-redshift
ULIGs (Sanders \& Mirabel 1996).

With the addition and refinement of data points along the SED, we can also fit
a dust temperature.  Fixing the dust emissivity at $\beta=1.5$, 
we find a best fit T$_{\rm d}=34$\,K, colder than the typical T$_{\rm d}$
found locally for sources extrapolated to this luminosity
(Chapman et al.~2002).
The implied gas-to-dust ratio is then $M(\hh)/M_{\rm d}\simeq$\,220.
This gas-to-dust ratio is at the lower end of the range of 
values seen in spiral galaxies (Devereux \& Young 1990), local ULIGs
(Sanders et al.~1991), and certainly other high-redshift CO sources.  
Hence the `Blob' seems
to represent a cooler system with different molecular
gas properties than other high-$z$ sub-mm galaxies, or local galaxies
with comparable far-IR luminosities.

\section{Discussion}

Having detected multiple irregular, and apparently
filamentary sources lying within the
Ly$\,\alpha$ cloud, it is tempting to attribute the 
Ly$\,\alpha$ emission to the ionizing photons from starbursts or AGN in these
objects.
The {\sl HST\/}-identified features are suggestive of
the first generation of merging between substantial fragments of galaxies.
We note that planar structures in the galaxy formation process,
or sequential star formation may also lead to such configurations ,
although the scales of the {\sl HST\/} fragments 
in the `Blob' are larger than any objects identified as `chain galaxies' in 
Cowie, Hu, Songaila (1995).
We further note that the radio-identified 
object, J1, which we identify with the sub-mm source,
lies in an apparent Ly\,$\alpha$ cavity in the integral field data of
Bower et al.~(2004), suggesting that huge winds may be driven by this
object.

Many pieces of evidence are consistent with this starburst scenario.
There are no obvious signatures of AGN, with non-detections in moderately
deep X-ray measurements, a weak radio source consistent with the
local far-IR--radio correlation (and thus suggestive of a starburst), 
and a rest-frame UV spectrum which
is inconsistent with an energetic AGN.

One problem with this scenario is that 
the equivalent width of the Ly$\,\alpha$ halo, under Case-B assumptions,
would imply a
difference in the apparent and required continuum luminosities of
a factor 2300. Even with very red spectra, it is impossible that
a correction to the star formation rates of the linear sources
could be this large (although for the most luminous infrared galaxies
seen locally, the star formation rates are factors of 100 larger than 
implied by the detectable UV emission -- Goldader et al.~2002).

One object which may be rather similar to the `Blob' is
SMMJ\,17142+5016, in the field of radio galaxy 53W002 at $z=2.4$.
This object is clearly identified as an AGN with an extended Ly$\,\alpha$
halo (Smail et al.~2003a). SMM\,J163650.0+405733 is also identified with 
an AGN, in the halo of extended Ly\,$\alpha$ and [OIII] (Smail et al.~2003b).
The lack of obvious signatures of AGN activity in the `Blob', and the clear 
presence of AGN in many other similar objects, suggest that we seriously
consider the possibility that an AGN is present in \ssa.

In this case, jet induced star formation may well be responsible for many of
the {\sl HST\/} fragments, as is likely the case in classical radio galaxies
(e.g.\ Pentericci et al.~1999).
If the extended NE structure (J4) represents emission from an 
AGN jet, the jet would then be roughly inclined at
$45^\circ$ to the plane of the sky, potentially also explaining
why the AGN itself is not seen directly in the X-ray
(its soft X-rays are mostly obscured by the dust torus).
This might also explain the general shape of
the huge Ly$\,\alpha$ reflection cloud, which might be interpreted loosely as a
triangular structure, in this case with a very wide opening angle
(${>}\,90^\circ$).  One might speculate even further about
a counter cloud emanating from J1 in the other direction
towards C11.  The scenario is similar to that outlined for the radio-loud
AGN, 53W002 (Windhorst et al.~1998), but on a much larger scale.
Deeper observations might indicate whether this picture is correct.

A likely scenario for the `Blob' is then a buried AGN. The {\sl Chandra\/}
non-detection in the
X-ray at first seems difficult to reconcile with the copious Ly$\,\alpha$
extended emission. However, it is much easier to obscure a compact 
X-ray emitting source that to preclude scattered photons from escaping
and ionizing the surrounding medium.
As with all submillimeter galaxies, the explicit presense or absense
of an AGN does not necessitate that it dominate the bolometric energy.
Ongoing star formation with cycles of AGN activity are the most likely 
scenario in the core of this massive proto-cluster region.
For now, we consider it a triumph that we have finally been able to pin
down the location of the submm and UV emission in this enigmatic object.

\section{Conclusions}

The \ssa\ object is still mysterious -- containing strong dust and Ly$\,\alpha$
emission coexisting in an environment that shows no unambiguous signs of AGN
activity.
A buried AGN may be the most likely explanation from energetic
grounds alone,
but this needs to be confirmed through some clear sign of AGN activity.
High ionization lines in the mid-IR
($\sim$15\mum) may be detectable in the `Blob' with {\sl SIRTF\/}-LWS if
excited by a dust-obscured AGN (see e.g., Rigopoulou et al.\ 1999).
It might still be that cooling radiation from a forming halo plays a role
(Fardal et al.\ 2001).
Further data will be required to determine whether multiple processes
contribute to the Ly$\,\alpha$ emission, and how these are related to the
sub-mm source.
Meanwhile it is worth investigating the relationship between Ly$\,\alpha$
and sub-mm emission for a wider sample of objects, to determine how unique
the \ssa\ object really is.

\acknowledgements
We thank A.\ Edge for access to the {\sl Chandra\/} X-ray data in this
region.
We gratefully acknowledge support from NASA through
{\sl HST\/} grant \#9174 (SCC, RW), 
awarded by the Space Telescope Science Institute. 
Based on observations made with the NASA/ESA Hubble Space Telescope,
obtained (from the Data Archive) at the
Space Telescope Science Institute, which is operated by the Association
of Universities for Research in Astronomy,
Inc., under NASA contract NAS 5-26555. 
The James Clerk Maxwell Telescope is operated by The Joint
Astronomy Centre on behalf of the Particle Physics and Astronomy Research
Council of the United Kingdom, the Netherlands Organisation for Scientific
Research, and the National Research Council of Canada.
CB and DS were supported by the Natural Sciences and Engineering Research
Council of Canada.

\end{document}